\documentstyle[aps,epsfig]{revtex}

\begin{document}

\wideabs{

\title{Motional narrowing effect in certain random binary lattices}

\author{F.\ Dom\'{\i}nguez-Adame}

\address{GISC, Departamento de F\'{\i}sica de Materiales, Universidad\\
Complutense, E-28040 Madrid, Spain}

\date{\today}

\maketitle

\tighten

\begin{abstract}

We present a model for a class of random binary lattices by introducing a
one-dimensional system where impurities are placed in one sublattice while host
atoms lie on the other sublattice. The source of disorder is the stochastic
fluctuation of the impurity energy from site to site. We study the optical
absorption spectra and the peculiarities of the motional narrowing effect at
the band edges for perturbative and nonperturbative degrees of disorder.
Analytical results agree well with numerical simulations.

\end{abstract}

}
\narrowtext

\section{Introduction}

Several years ago, Knapp raised the concept of motional or exchange narrowing
in one-dimensional (1D) disordered systems~\cite{Knapp84} and fruitfully
explained many optical phenomena in low-dimensional disordered systems like
J~aggregates and conjugated polymers (for a review, see
Refs.~\cite{Spano94,Knoester96} and references therein). Roughly speaking, this
author demonstrated that the disorder {\em seen\/} by the quasi-particles of
the excited system is reduced as compared to the seeding degree of disorder,
namely the width of the probability distribution of the on-site energy, as soon
as the states of the individual molecules are collectivized due to the
intermolecular interactions. The suppression factor depends on whether or not
the disorder is small enough to be regarded as a perturbation. In the
perturbative case, this factor is determined by the square root of the  number
of sites in the entire system, while in the nonperturbative case this number
should be substituted by the so-called number of coherently bound molecules,
due to the occurrence of the Anderson localization whenever the on-site energy
is an {\em uncorrelated\/} random variable. 

Numerically simulated absorption spectrum of poly\-silane with an uncorrelated
Gaussian  distribution of nearest-neighbor couplings are similar to those for 
an uncorrelated diagonal disorder~\cite{Tilgner90}. In contrast, simulations of
off-diagonal disorder given by Gaussian randomness in the molecular
positions~\cite{Fidder91} found that the behavior of the optical observables
does not fit the standard motional narrowing effect~\cite{Knapp84}. Recently,
this discrepancy has been uncovered by Malyshev and
Dom\'{\i}nguez-Adame~\cite{Malyshev99}, pointing out the appearance of
correlations in the intersite couplings even if the fluctuations of the
molecular positions are uncorrelated. Furthermore, it is nowadays well
established both theoretically~\cite{Flores89,Dunlap90} and
experimentally~\cite{Bellani99} that extended states may arise in 1D random
systems when disorder is correlated, in contrast to the earlier believe that
all eigenstates should be localized~\cite{Lee85}. The motional narrowing effect
and the competition between the localization length and the correlation length
have been considered very recently by Rodr\'{\i}guez {\em et
al.\/}~\cite{Rodriguez99}.

In this work, we report further progress along the lines in the preceding
paragraphs. In particular, we focus our attention on a new 1D tight-binding
model of correlated disordered system supporting extended
states~\cite{Hakobyan00,Adame00}. We built up our model by considering a 1D
random binary lattice with two species, referred to as A and B atoms hereafter.
We further assume that the site energy of A atoms is randomly distributed from
site to site while that of B atoms is the same over the other sublattice. We
have demonstrated analytically the occurrence of extended states in the
vicinity of the site energy of B atoms  {\em in spite of the fact that the
system is purely 1D and random\/}~\cite{Adame00}. Here we study the
peculiarities of the motional  narrowing effect and their manifestation through
the optical properties of this system.

\section{Model}

Let us consider a tight-binding Hamiltonian with nearest-neighbor interactions
\begin{equation}
{\cal H}=\sum_n\> \Big [ \epsilon_n |n\rangle\langle n|
-J |n\rangle\langle n+1| -J |n+1\rangle\langle n|\Big ],
\label{H}
\end{equation}
where the state vector $|n\rangle$ represents an excitation at site $n$. In the
present 1D binary system, A (B) atoms are placed at odd (even) positions of the
otherwise regular lattice, whose corresponding site energies are
$\epsilon_{2n-1}$ ($\epsilon_{2n}$) with $n=1,2,\ldots N$, $N$ being the number
of unit cells of the system. According to our model, site energies at even
positions are the same and we can set $\epsilon_{2n}=0$ without loss of
generality. The source of disorder arise from the stochastic fluctuations of
site energy at odd positions. We assume that $\{\epsilon_{2n-1}\}_{n=1}^{N}$ is
a set of {\em uncorrelated\/} random Gaussian variables with mean value $v$
and  variance $\sigma^2$. Hereafter $\sigma$ will be referred to as {\em degree
of disorder}. The joint distribution function is represented by the direct
product of single Gaussians. Thus 
\begin{equation}
\langle \epsilon_{2n-1} \rangle_{\mathrm av} = v, 
\quad \quad \langle \epsilon_{2n-1}\epsilon_{2n^{\prime}-1}
\rangle_{\mathrm av} = \left(v^2 + \sigma^2\right)\delta_{nn^{\prime}},  
\label{means}
\end{equation}
where the angular brackets $\langle \ldots \rangle_{\mathrm av}$ indicate the
average over the ensemble. Although the system is one-dimensional and random,
it has been demonstrated analytically the existence of a delocalized state in
infinite systems at $E=0$~\cite{Adame00}. Most important, there exist {\em
many\/} ($\sim \sqrt{N})$ states close to the resonant energy that remain
extended in finite systems, in the sense that their localization length is
larger than the system size.

\section{Perturbative motional narrowing effect}

We now calculate the motional narrowing effect in the perturbative limit by
considering small degree of disorder $\sigma$. To this end, we start by
writing down the eigenstates and eigenenergies of the unperturbed  Hamiltonian
($\sigma=0$), corresponding to a diatomic lattice with on-site energies $0$ and
$v$ in each unit cell. We assume rigid boundary conditions and $v>0$. The last
restriction is only for the sake of clarity in the exposition of results, as
they are trivially generalized to the case $v<0$. There are two allowed bands,
the first one ranging from $v/2-\sqrt{(v/2)^2+4J^2}$ up to $0$ and the second
one ranging from $v$ up to $v/2+\sqrt{(v/2)^2+4J^2}$ for $v>0$. The normalized
eigenstates of the unperturbed Hamiltonian can be written as
\begin{eqnarray}
|K,\pm\rangle&=&\sqrt{4\over 2N+1}\sum_{n=1}^{N}\>\Bigg\{\alpha_{K}^{\pm}
\sin[(2n-1)K]\,|2n-1\rangle \nonumber \\
&+&\beta_{K}^{\pm} \sin(2nK)\,|2n\rangle \Bigg\},
\label{eigenstate}
\end{eqnarray}
where $K\equiv k\pi/(2N+1)$ with $k=1,2,\ldots, N$ and
\begin{eqnarray}
\alpha_{K}^{\pm} &=& \left[ 1+{E_{K}^{\pm}\over E_{K}^{\pm}-v}\right]^{-1/2},
                 \nonumber \\
\beta_{K}^{\pm} &=& \left[ 1+{E_{K}^{\pm}-v\over E_{K}^{\pm}}\right]^{-1/2}.
\label{alpha_beta}
\end{eqnarray}
Here the labels $\pm$ refer to the upper ($+$) and the lower ($-$) band,
respectively. The dispersion relation within each band is given by
\begin{equation}
E_{K}^{\pm}=\frac{v}{2} \pm \sqrt{\frac{v^2}{4}+4J^2\cos^{2}K}.
\label{dispersion}
\end{equation} 

The perturbative part of the Hamiltonian, ${\cal H}_p$, is diagonal  in the
site representation and comes from the stochastic fluctuations of the on-site
energy at odd sites around the mean value $v$. Therefore ${\cal H}_p =
\sum_{n=1}^{N}\> (\epsilon_{2n-1}-v) |2n-1\rangle\langle 2n-1|$. In the
$K$-representation~(\ref{eigenstate}), the matrix elements of ${\cal H}_p$ are
expressed through linear combinations of Gaussian variables $\epsilon_{2n-1}-v$
with zero mean
\begin{eqnarray}
&&\langle K,\ell |{\cal H}_p |K^{\prime},\ell^{\prime} \rangle
={4\alpha_{K}^{\ell}\alpha_{K^{\prime}}^{\ell^{\prime}}\over 2N+1} \times
\nonumber\\
&&\sum_{n=1}^{N} \> (\epsilon_{2n-1}-v )\sin[(2n-1)K]\sin[(2n-1)K^{\prime}],
\label{matrix}
\end{eqnarray}
where $\ell$ and $\ell^{\prime}$ indicate the band. Consequently, they also
have a joint Gaussian distribution with zero mean. In what follows, we will
deal with the fluctuations of the square of the matrix elements given
in~(\ref{matrix}) since they determine the linear optical response of the
system.

Concerning optical transitions, since the states with $k=1$ (one per band)
carry almost the entire oscillator strength of the system, in the perturbative
limit the optical absorption spectrum is dominated by two Gaussian peaks
located at the bottom (top) of the lower (upper) band.  It is easy to
demonstrate that the oscillator strength is proportional to
$\alpha_{K}^{\pm}\mp \beta_{K}^{\pm}$, where $\alpha_{K}^{\pm}$ and
$\beta_{K}^{\pm}$ are given by~(\ref{alpha_beta}), so that the lower energy
transition becomes more intense. Therefore, we restrict ourselves to this
transition, whose standard deviation is given by $\sigma_1=B_{11}$ with
$B_{kk^{\prime}}^{2}\equiv \langle  \langle K,- |{\cal H}_p |K^{\prime},-
\rangle^{2}\rangle_{\mathrm av}$. After performing the
summations~(\ref{matrix}) one can obtain $B_{kk^{\prime}}$. Close to the zone
center, which is responsible of the linear optical response of the system, the
calculation yields 
\begin{equation}
B_{kk^{\prime}}=\frac{\sqrt{v^2+16J^2}-v}{2\sqrt{v^2+16J^2}}\,
\sqrt{2+\delta_{kk^{\prime}}\over 2N+1}\,\sigma, \quad k,k^{\prime} \ll N,
\label{B}
\end{equation}
where in~(\ref{alpha_beta}) we have replaced $E_{K}^{-}$ and
$E_{K^{\prime}}^{-}$ by their values at the zone center since $K\to 0$.  
Therefore, the standard deviation of the main absorption line is given by
\begin{equation}
\sigma_{1}=\frac{\sqrt{v^2+16J^2}-v}{2\sqrt{v^2+16J^2}}\,
\sqrt{3\over 2N+1}\,\sigma.
\label{sigma}
\end{equation}
As it can be seen from (\ref{sigma}), the standard deviation of the Gaussian
peak scales as $N^{-1/2}$, showing the so-called motional narrowing
effect~\cite{Knapp84}. Notice that the reduction factor carries information on
the system parameter $v/J$.

\section{Nonperturbative degree of disorder}

The perturbative approach we have carried out in the preceding section holds
for small degree of disorder. On increasing the degree of disorder, mixing of
the states $k=1$ with the other states strongly affect the system optical
response. Hence, one should compare the energy difference  between the two
lowest states, $\delta E_{12} \equiv |E_{K_{1}}^{-} - E_{K_{2}}^{-}|$, with the
typical fluctuation of $\langle K_{1},- |{\cal H}_p |K_{2},-\rangle$ 
represented by $B_{12}$, where $K_{m}\equiv m\pi/(2N+1)$. The perturbative
approach is valid provided $\delta E_{12} > B_{12}$ and fails otherwise. The
energy difference can be readily determined from~(\ref{dispersion})  
\begin{equation}
\delta E_{12} = \frac{12\pi^2J^2}{\sqrt{v^2+16J^2}}\,\frac{1}{(2N+1)^2},
\quad N \gg 1.
\label{delta}
\end{equation}
For higher values of the degree of disorder ($\delta E_{12}<B_{12}$) the
perturbative approach fails; not all sites contribute to the optical spectrum 
since the Anderson localization length becomes smaller than the system size.
Thus, $N$ should be replaced by $N^{*}<N$ in the above
equations~\cite{Knapp84}, where $N^{*}$ is often referred to as the number of
coherently bound molecules. Malyshev proposed a self-consistent estimation of
$N^{*}$ by applying the perturbative criterion mentioned above to a  typical
localization segment of length $N^{*}$~\cite{Malyshev91}. This approach yields
excellent results for uncorrelated and pairwise correlated diagonal
disorder~\cite{Adame99}. Thus, we use the condition $B_{12}^{*} = \delta
E_{12}^{*}$, where the asterisk means the substitution of $N$ by $N^{*}$
in~(\ref{B}) and~(\ref{delta}). In doing so we obtain
\begin{equation}
\frac{1}{\sqrt{2N^{*}+1}}=
\left(\frac{\sqrt{v^2+16J^2}-v}{12\sqrt{2}\pi^2J^2}\,\sigma\right)^{1/3}.
\label{N*}
\end{equation}
Finally, replacing in~(\ref{sigma}) $N$ by $N^{*}$ given by~(\ref{N*}) we
get for the standard deviation of the peaks
\begin{mathletters}
\begin{equation}
\sigma_{1}^{*}=C\,\frac{\sigma^{4/3}}{J^{1/3}},
\label{sigma*}
\end{equation}
where
\begin{equation}
C=\,\frac{\sqrt{3}}{2(12\sqrt{2}\pi^2)^{1/3}}
\,\frac{\left[\sqrt{(v/J)^2+16}-v/J\right]^{4/3}}{\sqrt{(v/J)^2+16}}.
\label{C}
\end{equation}
\end{mathletters}
This expression holds for $v>0$ as well as for $v<0$. The power-like dependence
$\sigma^{4/3}$  also appears in dealing with uncorrelated and pairwise
correlated diagonal disorder but with different $C$ (see, e.g.,
Ref~\cite{Adame99}) and can be obtained from a more rigorous framework using
the coherent potential approximation~\cite{Boukahil90}.  

\section{Numerical simulations}

\begin{figure}
\centerline{\epsfig{file=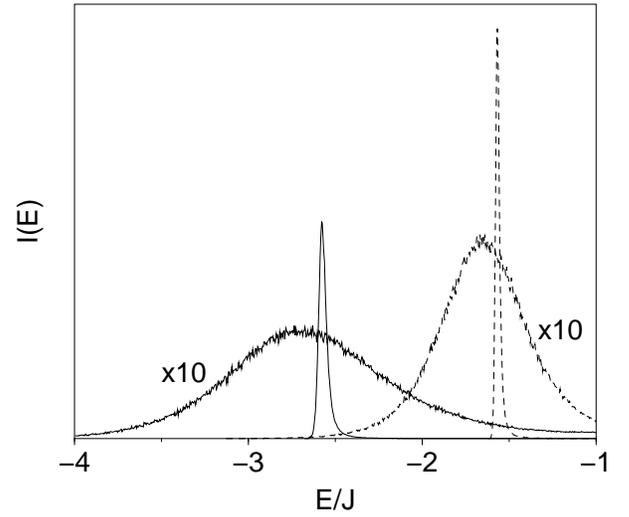,width=8cm}}
\caption{Absorption spectra for one-dimensional random binary lattices with
$J=1$, $N=125$ ($250$~sites) and $v=-1$ (solid lines) and $v=+1$ (dashed
lines). The narrow and broad peaks correspond to $\sigma=0.1$ and $\sigma=0.9$,
respectively. Notice the scale factor in the later case. Each spectrum
comprises the results of $10\,000$ realizations of the disordered system.}
\label{fig1}
\end{figure}

To check the accuracy of the above analytical approach, we have obtained
numerically the absorption line shape, $I(E)$,  according to 
Ref.~\cite{Fidder91}. We fix the value $J=1$ and focus our attention on the
degree of disorder $\sigma$, ranging from $0.1$ up to $0.9$. We have
diagonalized the Hamiltonian (\ref{H}) for chains of $N=125$ ($250$~sites) with
rigid boundary conditions. The number of randomly generated systems is
$10\,000$ for each value of $\sigma$. We show in Fig.~\ref{fig1} several
examples of the optical absorption spectra for various values of the parameters
$v$ and $\sigma$; only the lower energy side of the spectra is shown. For small
values of $\sigma$ the position of the main line agrees very well with the
value $v/2-\sqrt{(v/2)^2+4J^2}$ obtained from~(\ref{dispersion}). This main
line broadens and becomes redshifted on increasing the degree of disorder, as
expected.

\begin{figure} 
\centerline{\epsfig{file=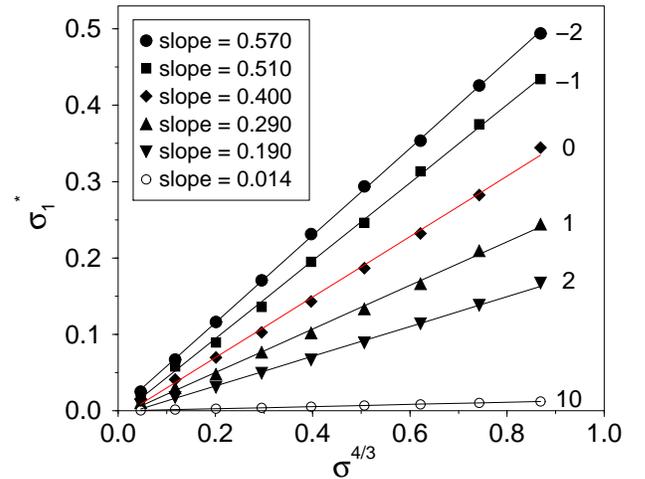,width=8cm}}
\caption{Standard deviation $\sigma_{1}^{*}$ as a function of $\sigma^{4/3}$ 
for different values of $v$, indicated at the right of each line, with $J=1$ and
$N=125$ ($250$~sites). Solid lines represent the least square fits and each
point comprises the results of $10\,000$ realizations of the disordered
system.} 
\label{fig2} 
\end{figure}

The value of $\sigma_{1}^{*}$ was obtained by nonlinear Gaussian fitting of 
the low energy side of the spectra and the results are presented in
Fig.~\ref{fig2}. It is found that this magnitude scales as $\sigma^{4/3}$  for
the various values of $v$ studied, as predicted by our
estimates~(\ref{sigma*}). The slopes of the straight lines are slightly larger
than those obtained from~(\ref{C}), as shown in Table~\ref{tab1}; nevertheless,
the coincidence should be admitted as being highly surprising in view of the
simple assumptions we made. 

\section{Conclusions}

In this paper we have considered the motional narrowing effect in a 1D random
binary lattice where disorder lies in one of the two sublattices. This system
could be regarded as a simple model of semiconductor binary alloy ---like
ternary III-V compounds---. In these alloys (say Al$_{x}$Ga$_{1-x}$As), the
cation sublattice is occupied by the same atoms (say As) while anions (say Al
and Ga) are randomly distributed over the other sublattice. Starting from a
perturbative approach, we found that the width of the absorption line scales as
$\sim N^{-1/2}$ for small degree of disorder, showing the motional narrowing
effect. For larger degrees of disorder, we determined self-consistently the
spatial extend of the excitation wave function, according to the prescription
raised by Malyshev~\cite{Malyshev91}. The spatial extend of the excitation wave
function is smaller than the system size and the width of the absorption line
scales as $\sim \sigma^{4/3}$. Our estimates agree well with numerically
simulated optical absorption spectra.

\acknowledgments 

The author thanks V.\ Malyshev and A.\ Rodr\'{\i}guez for helpful discussions.
This work is supported by Comunidad de Madrid under Project~07N/0034/98.

\begin{table}
\begin{center}
\begin{tabular}{|c|c|c|c|c|c|c|} 
$v/J$      &  -2  &  -1  &   0  &   1  &   2   &   10  \\ \hline\hline
Analytical & 0.42 & 0.34 & 0.25 & 0.17 & 0.12  & 0.010 \\ \hline
Numerical  & 0.57 & 0.51 & 0.40 & 0.29 & 0.19  & 0.014 \\ 
\end{tabular}
\end{center}
\caption{Slope of $\sigma_{1}^{*}$--$\sigma^{4/3}$ plots obtained analytically
from~(\protect{\ref{C}}) and numerically for different values of the parameter
$v$.}
\label{tab1}
\end{table}

\end{document}